\documentclass[preprintnumbers,amsmath,amssymb,prd,twocolumn,showpacs]{revtex4}
\usepackage{amsmath,amssymb,comment}

\renewcommand{\vec}[1]{\boldsymbol{\mathrm{#1}}}

\begin{document}

\title{Accelerating classical charges and the equivalence principle}


\author{Viktor T. Toth$^{\dag}$\\~\\}

\affiliation{$^\dag$Ottawa, Ontario K1N 9H5, Canada}



\begin{abstract}
We compare the behavior of a charged particle in a gravitational field and empty space. We resolve the apparent conflict between the Lorentz-Dirac equation and Larmor's formula of radiation by noting that the former describes an electron that is itself accelerated by an electromagnetic field. If instead, a hypothetical particle is considered that is accelerated by a non-electromagnetic force, Larmor's formula is found to be consistent with the accelerating particle's equation of motion. We consider the consequences concerning the equivalence principle and find that it is indeed violated if one demands that the same electromagnetic field be present in both the gravitational and accelerating cases; however, if one allows for the external electromagnetic fields to be different, the validity of the equivalence principle is restored. In either case, the basic idea behind the equivalence principle, which leads to a geometrized theory of gravity, remains unaffected.
\end{abstract}


\maketitle

Einstein's celebrated {\em gedankenexperiment} about an observer inside a windowless elevator cab not being able to distinguish being at rest in a gravitational field from accelerated motion in empty space has preoccupied many researchers over the past century.

A particular variation on this theme is when the elevator cab contains a charged particle. It is well established that an accelerating charge radiates electromagnetic energy \cite{1938RSPSA.167..148D,Rohrlich:1097602,Lyle:1339224,1980AnPhy.124..169B,1993gr.qc.....3025P}. On the other hand, a charge that is stationary in a gravitational field is not supposed to radiate; indeed, if it did, it could serve as a limitless source of radiative energy, a kind of {\em perpetuum mobile}.

That this is not the case can be shown in a variety of ways, and this result prompted some researchers to declare the principle of equivalence invalid. Clearly, the argument goes, the observer can tell easily if the elevator cab, pushed by a rocket engine, is accelerating in Minkowski space or is held still in a Schwarzschild spacetime. All that is needed is measuring the output of the rocket engine and comparing it to the readings of an accelerometer. To achieve the same acceleration in empty space, slightly more rocket power would be required, to account for the radiative energy loss due to the accelerating charged particle on board. Or to be more specific, by measuring the force required to hold the charged particle in place one can determine if a non-gravitational force is acting on the particle due to its accelerating motion.

But is this really a violation of the equivalence principle, or merely its misguided application? In the present paper, we argue for the latter interpretation.

The equation of motion of a free particle in curved spacetime is the geodesic equation:
\begin{align}
\frac{du^\mu}{d\tau}+\Gamma^\mu_{\kappa\lambda}u^\kappa u^\lambda=0,
\end{align}
where $u^\mu=dq^\mu/d\tau$ is the particle's 4-velocity, $q^\mu$ is its position, $\tau$ is proper time and $\Gamma$ represents the usual Christoffel-symbols.

In the presence of an external force, the particle no longer follows a geodesic. Its equation of motion becomes:
\begin{align}
\frac{du^\mu}{d\tau}+\Gamma^\mu_{\kappa\lambda}u^\kappa u^\lambda-a_{\rm ext}^\mu=0,
\end{align}
where $a_{\rm ext}^\mu=m^{-1} F_{\rm ext}^\mu$ represents the 4-acceleration due to the external force acting on a particle with mass $m$. For simplicity, in the following we use units such that $m=1$ and we characterize the external force by $a_{\rm ext}^\mu$.

On the other hand, if a particle has an electric charge $e$, its equation of motion in the presence of an electromagnetic field $F_{\mu\nu}$ is given by
\begin{align}
\frac{du^\mu}{d\tau}+\Gamma^\mu_{\kappa\lambda}u^\kappa u^\lambda=eF^\mu_\nu u^\nu.
\end{align}
For a particle that is influenced by both an external force and the interaction between its charge and the electromagnetic field, the equation of motion reads
\begin{align}
\frac{du^\mu}{d\tau}+\Gamma^\mu_{\kappa\lambda}u^\kappa u^\lambda-a_{\rm ext}^\mu=eF^\mu{}_\nu u^\nu.
\end{align}

Clearly in this form, the equation of motion can account for arbitrary trajectories in the presence of arbitrary electromagnetic fields, so long as an appropriate external force (characterized by $a_{\rm ext}^\mu$) is applied.

In the general case, the electromagnetic field will be the sum of any externally sourced fields and the ``radiation reaction'' field of the accelerating particle itself. However, for the sake of convenience, we may lump together the forces due to external fields and any non-electromagnetic forces into $a_{\rm ext}^\mu$. Under this definition, $F^\mu{}_\nu$ will represent the portion of the electromagnetic field that is due entirely to the particle itself. In other words, it will be determined by the 4-current
\begin{align}
J_\nu=\nabla_\mu F^\mu{}_{\nu},\label{eq:4current}
\end{align}
which will be zero everywhere except the particle's world line (here, $\nabla_\mu$ represents the covariant derivative). The covariant form of $J_\nu$ is given by \cite{1938RSPSA.167..148D}:
\begin{align}
J^\mu(x^\kappa)=e\int d\tau u^\mu\delta^{(4)}\big(x^\kappa-q^\kappa(\tau)\big),
\end{align}
where the particle's worldline, $q^\kappa(\tau)$, is parameterized by proper time $\tau$, serving as an affine parameter.

Solutions to (\ref{eq:4current}) can be deduced from the Lorentz-Dirac equation, which yields the following result for the electromagnetic field that corresponds to the radiation reaction force of the accelerating particle:
\begin{align}
F^\mu{}_\nu u^\nu=\frac{2e}{3}\left(\frac{d^2 u^\mu}{d\tau^2}+g_{\kappa\lambda}\frac{du^\kappa}{d\tau}\frac{du^\lambda}{d\tau}u^\mu\right).\label{eq:F}
\end{align}

For a particle at rest in the Schwarzschild metric, $u^\mu=[(1-2M/q^r)^{-1/2},0,0,0]$ and $du^\mu/d\tau=0$. The external force required to maintain the particle's position is determined by the equation of motion:
\begin{align}
\Gamma^\mu_{\kappa\lambda}u^\kappa u^\lambda-a_{\rm ext}^\mu=0,
\end{align}
which yields the standard Newtonian result, $a^r_{\rm ext}=M/r^2$ and all other components of $a^\mu_{\rm ext}$ being zero.

In contrast, in Minkowski spacetime with coordinates $(t,r,y,z)$, a particle uniformly accelerating in the $r$ direction follows the worldline given by $q^t=\alpha^{-1}\sinh \alpha\tau,$ $q^r=\alpha^{-1}\cosh \alpha\tau$, $q^y=q^z=0$. This trajectory yields $F^\mu{}_\nu u^\nu=0$ in (\ref{eq:F}), suggesting that the particle does not emit or absorb radiative energy. However, this is in direct contradiction with Larmor's well-known formula, which yields the radiative power of an accelerating charge as
\begin{align}
P=\frac{2e^2}{3}g_{\mu\nu}\frac{du^\mu}{d\tau}\frac{du^\nu}{d\tau}.
\end{align}

To resolve this conundrum, we first note that Eq.~(\ref{eq:F}) can be written after a bit of trivial algebra (utilizing the fact that $g_{\kappa\lambda}u^\kappa u^\lambda=1$) in the form
\begin{align}
F^\mu{}_\nu u^\nu=\frac{2e}{3}g_{\nu\lambda}\left(\frac{d^2 u^\mu}{d\tau^2}u^\lambda-\frac{d^2u^\lambda}{d\tau^2}u^\mu\right)u^\nu.\label{eq:F2}
\end{align}
For the uniformly accelerating particle, the part in parenthesis in Eq.~(\ref{eq:F}) is zero, allowing us to rewrite (\ref{eq:F2}) as
\begin{align}
F^\mu{}_\nu u^\nu=&\frac{2e}{3}g_{\nu\lambda}\times\label{eq:F3}\\
&\left(
g_{\kappa\eta}\frac{du^\kappa}{d\tau}\frac{du^\eta}{d\tau}u^\mu
u^\lambda
-g_{\kappa\eta}\frac{du^\kappa}{d\tau}\frac{du^\eta}{d\tau}u^\lambda
u^\mu
\right)u^\nu.\nonumber
\end{align}
This result of course identically vanishes, but that is not really the point: what is important is to note the fact that this is really Larmor's formula, repeated twice, with opposite signs. How can this be?

Close scrutiny of Dirac's paper \cite{1938RSPSA.167..148D} reveals the culprit. This paper describes, for the first time, the correct relativistic equation of motion of an {\em electron}. Why the emphasis? Because an electron does not interact with its environment except through gravity and electromagnetism (and, of course, the weak force, but our topic is the {\em classical} electron, not the charged elementary fermion of the present-day standard model of particle physics.) Thus, there is no $a_{\rm ext}$.

This consideration figures implicitly in Dirac's choice to include both the retarded and the advanced Li\'enard-Wiechert potentials in the electron's equation of motion. In this case, it makes perfect sense that the energy that the electron radiates is, in turn, absorbed by the electron from the field that is required to accelerate it; hence, the net radiative power will be zero.

But this need not be the case if we consider a hypothetical charged point source that is accelerated by some means other than the electromagnetic field, e.g., by a small rocket engine. In this case, only the retarded potential needs to be considered. The resulting equation of motion for the {\em uniformly accelerating particle} in Minkowski space, therefore, will be in the form
\begin{align}
\frac{du^\mu}{d\tau}-a_{\rm ext}^\mu=\frac{2e^2}{3}g_{\kappa\eta}\frac{du^\kappa}{d\tau}\frac{du^\eta}{d\tau}u^\mu,
\end{align}
or, after substituting the hyperbolic trajectory:
\begin{align}
\frac{du^\mu}{d\tau}-a_{\rm ext}^\mu=-\frac{2e^2}{3}\alpha^2u^\mu,
\end{align}
which determines the nonzero components of $a_{\rm ext}$:
\begin{align}
a_{\rm ext}^t&=\alpha\left(\sinh\alpha\tau+\frac{2e^2}{3}\alpha\cosh\alpha\tau\right),\\
a_{\rm ext}^r&=\alpha\left(\cosh\alpha\tau+\frac{2e^2}{3}\alpha\sinh\alpha\tau\right).
\end{align}
In Rindler coordinates, 
the spacelike component of the acceleration vector is written as
\begin{align}
a_{\rm ext}^R&=\alpha+\frac{2}{3}e^2\alpha^2,
\end{align}
indicating that the observer in an enclosed elevator cab would indeed see his rocket operating at a higher level of thrust, the excess corresponding to the $(2/3)e^2\alpha^2$ term above.

This is the origin of the perceived violation of the equivalence principle. An observer sitting inside the elevator cab can unambiguously distinguish being at rest in a gravitational field vs. accelerating in Minkowski space by measuring the extra force required the accelerate a charged particle in the latter.

But is this really a true violation? It depends on how exactly the equivalence principle of formulated in the presence of charged particles and electromagnetic fields. One possible formulation may go like this: {\em An observer carrying a charge cannot distinguish between remaining at rest in a gravitational field or accelerating in Minkowski space, in the presence of the {\rm same} external electromagnetic field.} This form of the equivalence is clearly falsified. However, there is another possible formulation:

{\bf An observer carrying a charge cannot distinguish between remaining at rest in a gravitational field or accelerating in Minkowski space, {\em provided that in the latter case, the external electromagnetic field is supplemented by an additional term that cancels the radiation reaction force.}}

In other words, without prior knowledge about the external electromagnetic environment, carrying a charge will not enable an observer to distinguish the effects of gravity from those of acceleration.

Of course arguably, one may simply formulate the equivalence principle by specifically excluding any non-gravitational interactions in the first place, and just concentrate on the principle's {\em raison d'etre}: that it enables us to treat all massive objects the same regardless of their material composition, and thus craft a geometric theory of gravity.

\bibliography{refs}

\end{document}